\pdfoutput=1
\documentclass[aps,pra,twocolumn,floatfix,reprint]{revtex4-1}
\usepackage{amsfonts}
\usepackage{mathrsfs}
\usepackage{amsmath}
\usepackage{color} 
\usepackage{graphicx} 
\usepackage{bm}
\usepackage{amssymb} 
\usepackage[normalem]{ulem}
\usepackage{xspace} 
\usepackage{epstopdf} 
\usepackage{dcolumn}
\usepackage{tabularx} 
\usepackage{longtable} 
\usepackage[colorlinks=true, letterpaper=true, pdfstartview=FitV, linkcolor=blue, citecolor=blue, urlcolor=blue]{hyperref}
\newcommand{\sgn}{\text{sgn}}
\usepackage[normalem]{ulem}

\begin{document}
\title{Valley-Selective Topologically Ordered States in Irradiated Bilayer Graphene}
\author{Chunlei Qu}
\author{Chuanwei Zhang}
\author{Fan Zhang}
\email{zhang@utdallas.edu}
\affiliation{Department of Physics, University of Texas at Dallas, Richardson, TX 75080, USA}

\begin{abstract}
Gapless bilayer graphene is susceptible to a variety of spontaneously gapped states. 
As predicted by theory and observed by experiment, the ground state is however topologically trivial,
because a valley-independent gap is energetically favorable.
Here, we show that under the application of interlayer electric field and circularly polarized light, 
one valley can be selected to exhibit the original interaction instability while the other is frozen out. 
Tuning this Floquet system stabilizes multiple competing topologically ordered states, 
distinguishable by edge transport and circular dichroism. 
Notably, quantized charge, spin, and valley Hall conductivities coexist in one stabilized state.
\end{abstract}
\maketitle

\textcolor{blue}{\it{Introduction.---}}
Chirally stacked few-layer graphene, ranging from the Bernal bilayer to its thicker cousins with Rhombohedral stacking,
has become a paradigmatic platform for studying fascinating two-dimensional electron physics~\cite{Neto2009,Min-Review,Zhang2011,Qiao-Niu}. 
Unlike the linear Dirac bands in monolayer graphene, 
the bands in an $N$-layer system exhibit flatter dispersions $\sim\pm k^N$. 
Notably, the Fermi surface only consists of band touching points at two inequivalent hexagonal Brillouin zone corners,
known as $K$ and $K'$ valleys. Intriguingly, the Fermi points are protected by the quantized Berry phases $\pm N\pi$ 
in the presence of a chiral symmetry between the two sublattices located at the top and bottom layers.
This unique feature leads to band gap opening 
when an interlayer electric field breaks the chiral symmetry~\cite{McCann2006,Ohta2006,YBZ2009,Mak2009,Lui,Zhu}. 
More remarkably, due to the large density-of-states near the Fermi points, 
the $N\!>\!1$ systems are susceptible to a variety of broken chiral symmetry states~\cite{Zhang2015},
in which each spin-valley flavor spontaneously transfers charge between layers to yield opening of quasiparticle energy gaps~\cite{Min2008,Sun,Zhang2010,Levitov1} and spreading of momentum-space Berry curvature~\cite{Zhang2011}. 

The order competing is enriched by the $\mathcal{SU}(4)$ spin-valley symmetry\cite{Zhang2015,Min2008,Sun,Zhang2010,Levitov1,Vafek1,Falko1,Levitov2,Jung2011,Zhang2012,Honerkamp1,Honerkamp2,Trushin2011,Falko2,Vafek2,Yan,Gorbar,Min12,Kharitonov,Varma,Li2014,Rossi,ex1,ex2,ex,ex3,ex4,ex5,ex6,ex7,ex8,ex9,ex10,ex11}. 
The nearly-degenerate ground states can be topologically classified 
based on the signs of spontaneous gaps at each spin-valley~\cite{Zhang2011,Levitov2}, 
analogous to those single-particle states in Haldane and Kane-Mele models~\cite{Haldane1988,Kane2005}. 
The intervalley exchange interactions, although extremely weak and often negligible, energetically favor
those candidates with valley degeneracy~\cite{Jung2011}, 
i.e., the same layer polarization at the two valleys.
In order to minimize the electrostatic energy, hence the two spin flavors must polarize toward opposite layers .
These two facts yield a topologically trivial layer-antiferromagnetic (LAF) ground state~\cite{Zhang2011,Jung2011,Zhang2012,Kharitonov}, 
as recently confirmed by experiment~\cite{ex4,ex5,ex6,ex7,ex8}. 

\begin{figure}[b!]
\centering\includegraphics[width=0.45\textwidth]{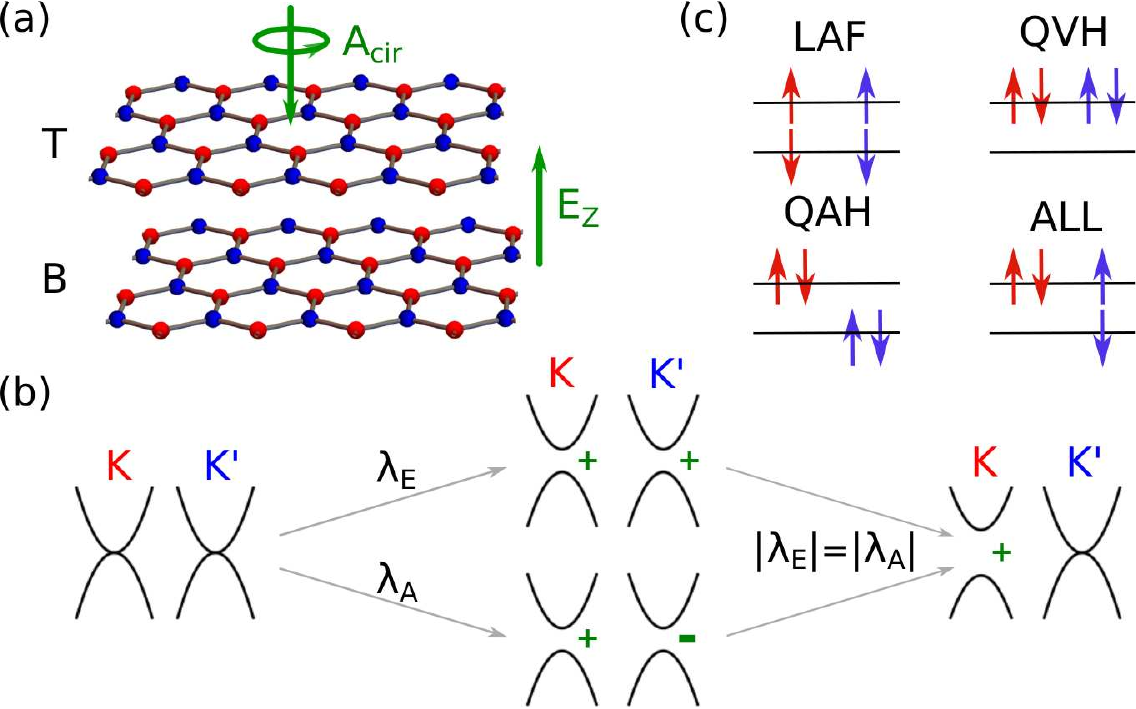}
\caption{(a) Sketch of bilayer graphene under the application of an interlayer electric field and a circularly polarized light. 
(b) The electric (light) field opens an energy gap with the same (opposite) sign(s) at the two valleys. 
If both fields generate equal gap magnitudes, the energy gap of a selected valley vanishes. 
(c) Illustration of the layer polarizations in the four competing ordered states. 
The up/down arrows denote the two spins; the red/blue colors denote the two valleys.}
\label{fig-bandstructure}
\end{figure}

One may naturally wonder whether and how the more exotic topologically ordered states~\cite{Zhang2011,Levitov2} can be stabilized. 
The key is to relax the aforementioned valley degeneracy; 
a possible route is to explicitly break both time reversal ($\mathcal{T}$) 
and spatial inversion ($\mathcal{P}$) symmetries that dictate the valley degeneracy.
We find that under the application of interlayer electric field 
and circularly polarized light in bilayer graphene, 
one valley can be selected to exhibit the original $\pm k^N$-band touching~\cite{Zhang2011} 
and weak interaction instability~\cite{Min2008,Sun,Zhang2010,Levitov1}, 
whereas the other valley is frozen out due to a large field-induced gap, 
as illustrated in Fig.~\ref{fig-bandstructure}. 
When the interaction-driven gap at the selected valley is opposite to the field-induced gap at the frozen valley, 
the ground state is a quantum anomalous Hall (QAH) state. 

Remarkably, when the selected-valley gaps become opposite between the
two spin flavors, quantized charge, spin, and valley Hall conductivities coexist. 
As we will examine, such an exotic ordered state dubbed ``ALL'' hereafter, 
elusive in the absence of the delicately applied fields or the electron-electron interactions,
can be stabilized by tuning the interacting Floquet system,
and different ordered states can be distinguished by edge transport and circular dichroism. 
We note that a light-irradiated Floquet state 
has been observed on a topological insulator (TI) surface~\cite{Wang2013,Galitski},
and our predicted ordered Floquet states may be similarly studied  
in chiral graphene experiments~\cite{ex1,ex2,ex,ex3,ex4,ex5,ex6,ex7,ex8,ex9,ex10,ex11}.

\textcolor{blue}{\it{Floquet theory.---}} 
We first establish our notation by discussing the effective Hamiltonian~\cite{Zhang2011}
that can coherently describe graphene ($N$$=$$1$), its Bernal bilayer ($N$$=$$2$), 
and its Rhombohedral few-layers ($N$$>$$2$),
to which our main results derived for the bilayer system can be generalized.
Such a model reads
\begin{equation}
h_{N} =\frac{(v_0\hbar k)^N}{(-\gamma_1)^{N-1}} [\cos(N\phi_{\bm{k}})\sigma_x
+\sin(N\phi_{\bm{k}})\sigma_y]+\lambda\sigma_z\,,
\label{HN}
\end{equation}
where $v_0$ is the Fermi velocity of graphene, $\cot\phi_{\bm{k}}=\tau k_x/k_y$, 
and $\tau\!=\!\pm$ label the $K$ and $K'$ valleys.
$\gamma_1\!\sim\!0.4$~eV is the interlayer nearest-neighbor hopping energy, 
which sets the largest energy scale of the model. 
The Pauli matrices $\bm{\sigma}$ act on the layer-pseudospin space 
spanned by the two sublattices relevant at low energy, 
i.e., the top $A$ and bottom $B$ without interlayer nearest neighbors. 
For $\lambda\!=\!0$, the gapless spectrum of $h_{N}$ has two band touch points at $K$ and $K'$,
protected by the chiral ($\mathcal{C}$) symmetry between the two sublattices.

$\lambda\sigma_z$ is a mass term that characterizes $\mathcal{C}$ symmetry breaking 
and hence opens an energy gap whenever $\lambda\neq 0$.
It turns out that the mass term can be explicitly induced by an interlayer electric field or a circularly polarized light, 
or spontaneously generated by electron-electron interactions.  
In the first scenario, as demonstrated by the experiments in $N\!>\!1$ systems~\cite{McCann2006,Ohta2006,YBZ2009,Mak2009,Lui,Zhu}, 
the interlayer electric field $E_z$ breaks $\mathcal{P}$ and $\mathcal{C}$ symmetries by producing
\begin{equation}
\lambda = \lambda_E = - e E_z d\,,
\label{UE}
\end{equation}
where $d$ is the layer separation. 
$\lambda_E$ is spin-valley independent, as required by $\mathcal{T}$ and spin $\mathcal{SU}(2)$ symmetries. 

To demonstrate the second scenario, consider a circularly polarized light shining on a few-layer.
This amounts to applying a time-dependent electromagnetic gauge potential
$\bm{A}(t)\!=\!A_0(\xi\sin\omega t, \cos\omega t)$ to the system,
where $\xi\!=\!\pm$ denote the light helicities. 
In Floquet theory, such a periodically driven system can effectively be described by a static Hamiltonian~\cite{Goldman2015,Bukov2015}
\begin{eqnarray}
h_{\rm eff}=h_{0}+\frac{1}{\hbar\omega}\sum_{j=1}^{\infty}\frac{1}{j}
\left[\mathcal{V}_{+j}, \mathcal{V}_{-j}\right]+\mathcal{O}\left(\frac{1}{\omega^2}\right)\,.
\label{FL}
\end{eqnarray}
$h_{0}$ is the time-independent Hamiltonian without the periodic drive; 
$\mathcal{V}_{\pm j}=(\omega/2\pi)\int \mathcal{V}(t) e^{\mp ij\omega t}dt$ 
are the Fourier components of the time-dependent periodic potential $\mathcal{V}(t)$. 
Similarly to how Eq.~(\ref{UE}) was derived~\cite{ZhangABC},
we apply $\bm{A}(t)$ to the original full-band model of $N$-layer graphene, 
followed by a projection to the two-band model at low energy. 
In the {\it high-frequency} limit, to the leading order the circularly polarized light yields
\begin{eqnarray}
\lambda=\lambda_{A}\,\tau = \frac{e^2v_0^2A_0^2}{\hbar\omega}\,\xi\,\tau\,.
\label{UA}
\end{eqnarray} 
The radiation field can yield a mass term $\lambda_A$ 
as its definite helicity breaks $\mathcal{T}$ and $\mathcal{C}$ symmetries.
The intact spin $\mathcal{SU}(2)$ and $\mathcal{P}$ symmetries dictate
$\lambda_A$ to be spin independent but opposite at the two valleys. 
$\lambda_A$ was originally derived in the seminal work on radiated monolayer graphene~\cite{Oka2009,Kitagawa2010,Morell2012,Piskunow2014},
and a similar gap has been experimentally observed on the TI surface~\cite{Wang2013,Galitski}.
Here we generalize the idea to graphene few-layers and demonstrate the general validity of Eq.~(\ref{UA}).

In a many-body scenario, Coulomb interactions can generate masses 
in the quasiparticle spectra of $N\!>\!1$ systems~\cite{Zhang2011}. 
While a microscopic theory will be given later, 
the ground state turns out to be LAF~\cite{ex5,ex6,ex7,ex8,ex9,ex10}, breaking all the symmetries. 
The two spin flavors are polarized to opposite layers spontaneously 
as if they are subjected to opposite $\lambda_E$ mean-fields,
i.e., $\lambda\!\propto\!s_z$ in Eq.~(\ref{HN}). 

In each scenario, the mass generation in Eq.~(\ref{HN}) spreads the Berry curvature, 
which is integrated to $N\sgn(\lambda)\tau/2$ for each spin-valley~\cite{Zhang2011}.
Hence, $\lambda_A$ induces a QAH state with Chern number $2N$,
which is reminiscent of the Haldane model~\cite{Haldane1988}; 
$\lambda_E$ produces a quantum valley Hall (QVH) state 
since the valley Chern numbers are opposite for different valleys~\cite{Zhang2011,vc1,vc2,vc3,vc4,vc5,vc6}, 
analogous to the quantum spin Hall state~\cite{Kane2005}. Metallic in-gap states have indeed been experimentally observed~\cite{vc7,vc9,vc10,vc11}
along a QVH domain wall in bilayer graphene where the edge conductance appears to approach $4e^2/h$.
By contrast, the LAF state is topologically trivial since it can be viewed as two opposite copies of QVH states. 
As motivated in {\it Introduction} and illustrated in Fig.~\ref{fig-bandstructure}(b), 
when $|\lambda_A|\!=|\lambda_E|$, 
one valley can be selected to exhibit the original $\pm k^N$-band touching, 
whereas the other valley is almost frozen out due to a large field-induced gap. 
The interactions generate masses at the selected valley,
whose signs determine three competing ordered states:
QAH, QVH, and the emergent ALL states.
With masses opposite (the same) in sign for the two spins at the selected (frozen) valley, 
the ALL state can be viewed as a QVH state for one spin but QAH for the other. 
Remarkably, such a state breaks all the symmetries and exhibits charge, spin, and valley Chern numbers of $N$. 
More remarkably, the ALL state is a synergistic consequence of intrinsic interactions and external fields, 
instead of due to any magnetic moment or spin-orbit coupling.

\textcolor{blue}{\it{Hartree-Fock theory $\&$ phase diagram.---}} 
We now study the phase diagram enrichment 
in terms of the following ordered state quasiparticle Hamiltonian~\cite{Zhang2011,ex5,ex6}:
\begin{subequations}
\begin{eqnarray}
\mathcal{H}^{\rm HF} &=& \sum\limits_{\bm{k}\alpha\beta s\tau} c_{\bm{k}\alpha s\tau}^\dagger
[h_N+h_H+h_F+h_V]c_{\bm{k}\beta s\tau} \,,\\
h_{H} &=& [V_0\Delta_0\delta^{\alpha\beta}+V_z\Delta_z\sigma_z^{\alpha\beta}]\,,\\
h_{F} &=&-[V_0+V_z\sigma_z^{\alpha\alpha}\sigma_z^{\beta\beta}]\Delta^{s\tau}_{\alpha\beta}\,,\\
h_{V} &=&-[V_0'+V_z'\sigma_z^{\alpha\alpha}\sigma_z^{\beta\beta}]\Delta^{s\bar\tau}_{\alpha\beta}\,,
\end{eqnarray}\label{Hhf}
\end{subequations}
which has well reproduced the experimentally observed gap size and $T_c$ in bilayer graphene.
In Eq.~(\ref{Hhf}), Greek letters label layer, $s$ labels spin, and $\tau$ labels valley. 
$V_{0,z}$$=$$(V_s\!\pm\!V_d)/2$ denotes the average (difference) of intralayer and interlayer interactions at the same valley, 
and likewise $V_{0,z}'$$=$$(V_s'\!\pm\!V_d')/2$ for valley-exchange interactions~\cite{Jung2011}.
The density matrix $\Delta^{s\tau}_{\alpha\beta}$$=$$\sum_{\bm{k}}\langle c_{\bm{k}\alpha s\tau}^\dagger c_{\bm{k}\beta s\tau} \rangle_f/A$ 
must be determined self-consistently. We introduce $\Delta_{0,z}^{s\tau}$ 
as the density sum (difference) of the top and bottom layers for one spin-valley flavor
and $\Delta_{0,z}$$=$$\sum_{s\tau}\Delta_{0,z}^{s\tau}$ as the total density sum (difference).
The mean-field interaction vertices must be diagonal in layer due to in-plane rotational symmetry;
neither spin nor valley coherence can be established spontaneously 
since the long-range Coulomb interactions $V_{0,z}$ dominate.
Therefore, in Eq.~(\ref{HN}) $\lambda$ reads
\begin{equation}
\lambda^{s\tau}=\lambda_{E} + \lambda_{A}\tau + V_z\Delta_z-\frac{V_s}{2}\Delta_z^{s\tau}-\frac{V_s'}{2}\Delta_z^{s\bar{\tau}}\,,
\label{Gap}
\end{equation}
where $\Delta_{z}^{s\tau}$ minimize the energy density functional
\begin{eqnarray}
\!\!\!\!\varepsilon_g=&-&\frac{1}{A}\sum_{\bm{k}s\tau}\sqrt{(\lambda^{s\tau})^2+(v_0\hbar k)^{2N}/\gamma_1^{2N-2}} \nonumber \\
&-&\bigg[\frac{V_z}{2}(\Delta_z)^2 \!-\!\frac{V_s}{4}\sum_{s\tau}(\Delta_z^{s\tau})^2 \!-\!\frac{V_s'}{4}\sum_{s\tau}\Delta_z^{s\tau}\Delta_z^{s\bar{\tau}}\bigg]\,.
\label{Eg}
\end{eqnarray}

Given that the subtracted in the second line of Eq.~(\ref{Eg}) is 
exactly half of the mean-field interactions that are implicit in the first line,
the three interaction parameters ($V_s\!\gg\!V_z,\!V_s'\!>\!0$) play different roles in order competing.
The intralayer exchange $V_s$ causes spontaneous layer polarization in each spin-valley,
whereas the Hartree energy determined by $V_z$ prevents any total layer polarization.
Although rather weak, the valley-exchange interaction $V_s'$ favors a ground state 
in which different valleys have the same layer polarization. 
Therefore, for $E_z\!=\!A_0\!=\!0$, the ground state must be the LAF state 
in which different spin (valley) flavors are layer polarized in the opposite (same) sense,
as observed in experiment~\cite{ex5,ex6,ex7,ex8,ex9,ex10}.

\begin{figure}[t!]
\centering\includegraphics[width=0.48\textwidth]{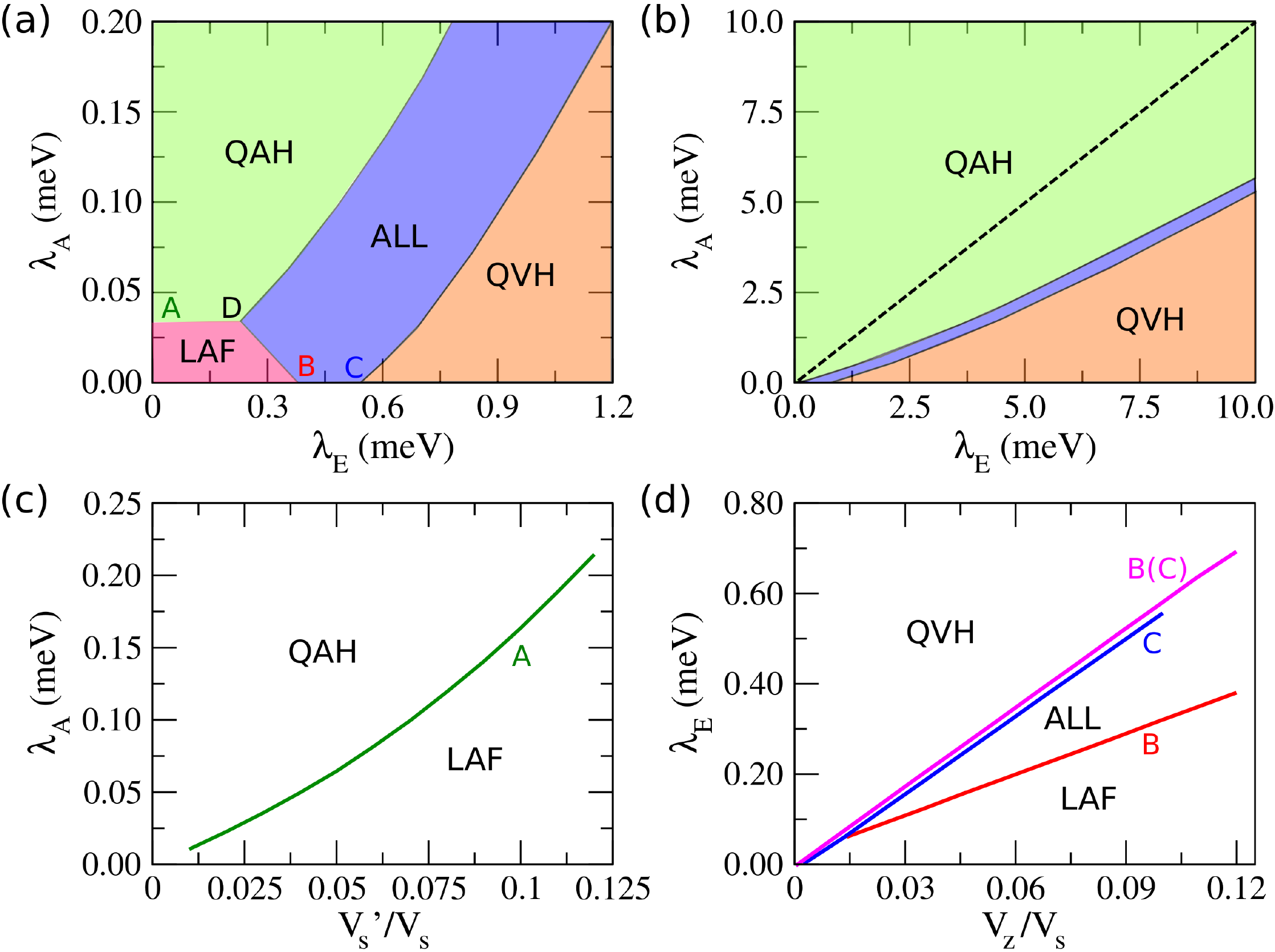}
\caption{(a)-(b) Phase diagram of bilayer graphene at small and large fields, respectively.
The Hartree interaction $V_z\!=\!0.1V_s$; the valley-exchange interaction $V_s'\!=\!0.03V_s$.
(c)-(d) The dependences of the critical points $A$-$C$ in (a) on  $V_s'$ and on $V_z$, respectively.
The red and blue lines are for $V_s'\!=\!0.02V_s$, and the purple line is for $V_s'\!=\!0.08V_s$.}
\label{fig-diagram}
\end{figure}

We now examine how the circularly polarized radiation enriches the order competing 
and stabilizes the ordered QAH and ALL states.
We focus on the $N$$=$$2$ case to facilitate our numerics; 
the qualitative results should apply to $N$$>$$2$ cases.
Fig.~\ref{fig-diagram}(a) shows the phase diagram, 
which exhibits four competing orders and mirror symmetries with respect to $\lambda_{E,A}\!=\!0$ lines. 
The LAF state exists in the limit of vanishing external fields.
A sufficiently large $\lambda_E$ ($\lambda_A$) favors the QVH (QAH) state.
Near the $|\lambda_E|\!=\!|\lambda_A|$ line, there emerges the ALL state.

In the phase diagram Fig.~\ref{fig-diagram}(a), the positions of the critical points $A$-$D$ 
are determined by the strengths of three interaction parameters.
This has also been discussed in the above analysis of Eqs.~(\ref{Gap})-(\ref{Eg}).
(i) The experimental LAF gap is about $2$~meV at zero fields~\cite{ex5,ex6}; 
it follows from Eqs.~(\ref{Gap})-(\ref{Eg}) that the dominating intralayer exchange $\nu_0 V_s\!=\!0.2992$,
where $\nu_0$ is the density of states per flavor.
(ii) Although both QAH and LAF states are ordered without total layer polarization, 
it is the weak valley-exchange interaction $V_s'$ that favors the LAF state~\cite{Jung2011}, 
in which different valleys have the same layer polarization. 
Consequently, the $A$ point shifts toward a larger $\lambda_A$ if $V_s'/V_s$ is stronger, 
as shown in Fig.~\ref{fig-diagram}(c).
(iii) The weak Hartree interaction $V_z$ prevents any total layer polarization induced by finite $\lambda_E$.
Thus, both $B$ and $C$ shift toward larger $\lambda_E$'s if $V_z/V_s$ is stronger, 
as shown in Fig.~\ref{fig-diagram}(d).
(iv) Interactions favor the QAH state over the QVH state and produces the novel ALL state near their transition. 
The ALL-state phase boundaries in Fig.~\ref{fig-diagram}(b) are shifted away from the $|\lambda_E|\!=\!|\lambda_A|$ line, 
which separates the QAH and QVH states in the absence of interactions. 
Such a shift arises from the compensation of the Hartree energy by the $\lambda_E$ mass in the frozen valley $\tau$, 
and it is given by $|\lambda_E|\!-\!|\lambda_A|\!\sim\!V_z|\Delta_z^{\tau}|$.
The width of the ALL-state regime is similarly determined, but by the selected valley ${\bar\tau}$,
and it is given by $\delta|\lambda_E|\!\sim\!V_z|\Delta_z^{\bar\tau}|$.
Note that since $B$ is sensitive to the change of $V_s'$ whereas $C$ is not, 
the ALL state may disappear at $\lambda_A\!=\!0$ if $V_s'$ exceeds a critical value,
as exemplified by the magenta line in Fig.~\ref{fig-diagram}(d). 
(Here we predict that a circularly polarized light can stabilize and control the ALL state without Landau levels.
Its quantum Hall ferromagnetic variant has been experimentally observed at $\nu\!=\!2$~\cite{ex9}.)

The parameters for the emergence of the intriguing ALL state in the phase diagram Fig.~\ref{fig-diagram}(b)
can be fitted to $\lambda_A\approx 0.6\lambda_E - 1$~meV at relatively large fields. 
To observe the ALL state, e.g., one can scan the parameter region around $\lambda_E\sim 10$~meV and $\lambda_A\sim 5$~meV. 
Given Eq.~(\ref{UE}), the electric field strength is $E_z=\lambda_E/(ed)\sim 30$~mV$/$nm where $d=3.4$~\AA~is the interlayer separation. 
The laser frequency, which should be much larger than the interaction-driven or field-induced energy gap $\sim 10$~meV 
but much smaller than $\gamma_1\sim 0.4$~eV, can be chosen as $\hbar\omega\sim 100$~meV. 
This corresponds to a light frequency $\sim 25$~THz or a light wavelength $\sim 12$~$\mu$m. 
Given Eq.~(\ref{UA}), we can obtain the energy flux of the light is $I=\omega^2 A_0^2/\mu_0 c\sim 3\times 10^{10}$~W$/$m$^2$.

\begin{figure}[t!]
\centering\includegraphics[width=0.48\textwidth]{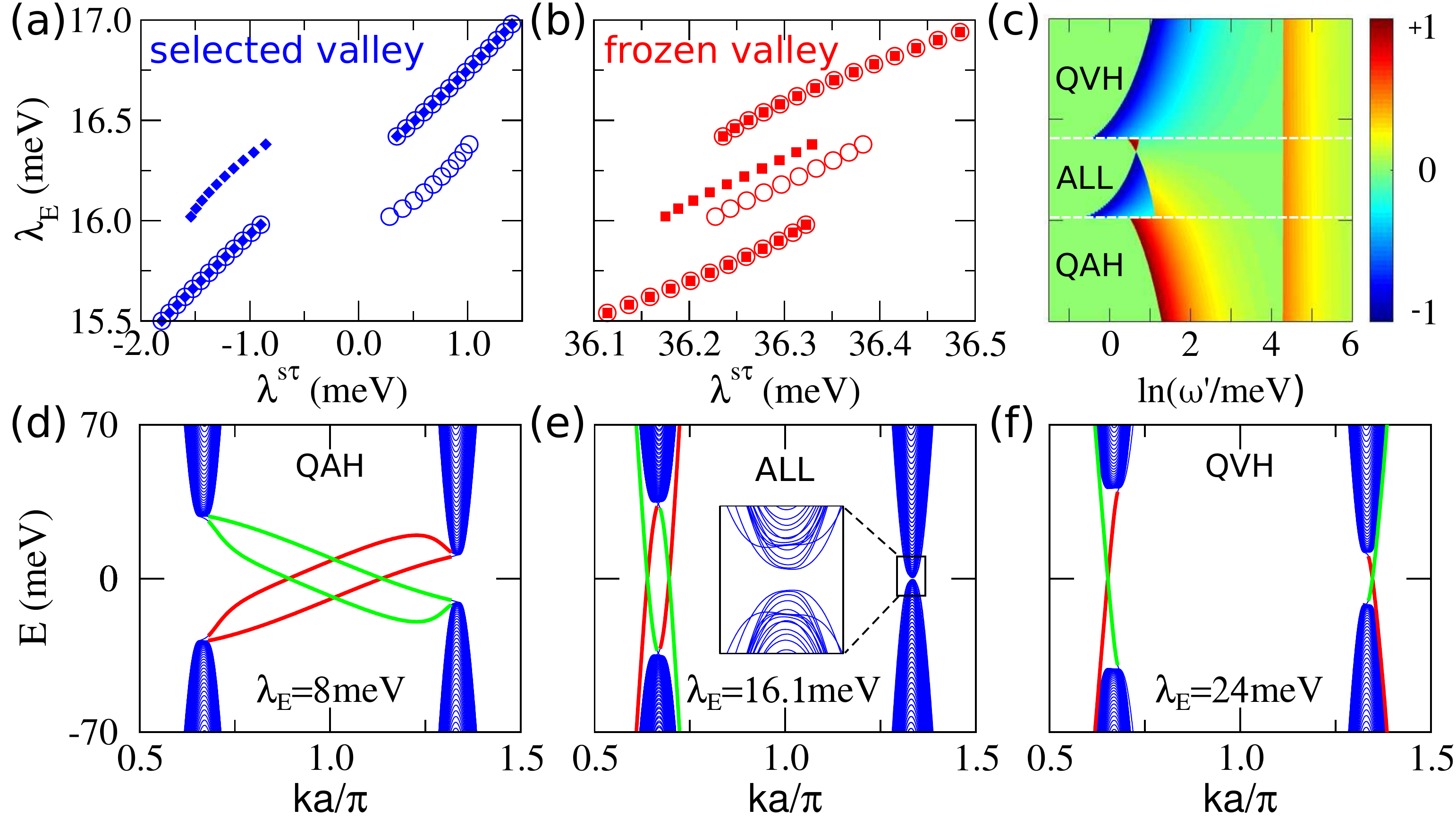}
\caption{(a)-(b) Ordered state masses $\lambda^{s\tau}$ and (c) circular dichroism $\eta(\omega')$ of bilayer graphene,
as functions of $\lambda_E$.  The square and circle denote different spins,
$\lambda_A$ is fixed at $10$~meV, and other parameter values are the same as in Fig.~\ref{fig-diagram}(b).
(d)-(f) Protected zigzag edge states for the three phases. The red and green states, 
spin degenerate in (d) and (f) but non degenerate in (e), localize at opposite edges.
Note that a zigzag edge is the best to observe the charateristic edge states but not required. 
The edge states are robust as long as the two valleys are well separated in the edge Brillouin zone 
and the boundary (or domain wall) is smooth in a scale much larger than the inverse of the separation.}
\label{fig-probe}
\end{figure}

\textcolor{blue}{\it{Circular dichroism.---}}
In addition to the number of valley-projected chiral edge states [Fig.~\ref{fig-probe}(d)-(e)] dictated by the aforementioned Chern numbers, 
different topologically ordered states may also be characterized by optical means. 
Here we consider {\it terahertz} absorbance~\cite{ab1,ab2} and its dichroism~\cite{Probe1,Probe2,Probe3,Probe4} 
using a {\it second}, circularly polarized, normally incident light. 
Consider a {\it weak} probe beam $\bm{A}'\!=\!A_0'(\xi'\sin\omega' t,\cos\omega' t)$,
and to the first order the induced perturbation reads
\begin{equation}
\mathcal{V}'=\frac{ie\hbar v_0^2 A_0'}{\gamma_1}(\tau k_x-ik_y)(\sigma_x + i\sigma_y)e^{i\xi'\tau\omega' t} + \text{h.c.}\,.
\end{equation}
Using Fermi's golden rule, we obtain the interband transition probability,
followed by the flavor absorbance
\begin{equation}
P_{\xi'}^{s\tau}=\frac{\pi\alpha}{2}\left(1+ \xi'\tau \frac{2\lambda^{s\tau}}{\hbar\omega'} \right)^2 \Theta(\hbar\omega'-2|\lambda^{s\tau}|)\,,
\end{equation}
where $\alpha$ is the fine structure constant. 
We further define the total absorbance and the circular dichroism as
\begin{equation}
P_{\xi'}=\sum_{s\tau}P_{\xi'}^{s\tau} \quad \mbox{and}\quad
\eta = \frac{P_+-P_-}{P_++P_-}\,.
\end{equation}
In the limit of $\hbar\omega'\!\gg\!|\lambda|$ or $\lambda\!\rightarrow\!0$, 
the total absorbance recovers the universal result $2\pi\alpha$,
independent of the light helicity or polarization. This results in $\eta\!\rightarrow\!0$.
Close to the thresholds, $\hbar\omega'=2|\lambda^{s\tau}|$,  
a circularly polarized light is either destructively blocked or constructively absorbed,
depending on the light helicity, the valley index, and the mass sign. 
This leads to sharp peaks in $\eta$.

Figure~\ref{fig-probe} plots $\lambda^{s\tau}$ and $\eta(\omega')$ as functions of $\lambda_E$
for the case of $\lambda_A\!=\!10$~meV.
As the electric field increases, the ground state undergoes QAH-ALL-QVH transitions. 
At the frozen valley the masses are large and of uniform sign for the two spins,
whereas at the selected valley the masses are small and relatively inverted for 
zero, one, and two spin flavors, respectively, in QVH, ALL, and QAH states.
Universally, $\eta$ shows a marked jump when the selected valley starts to be probed
and gradually falls upon increasing $\omega'$, except that $\eta$ switches sign for the ALL state 
due to the different magnitudes of $\lambda^{s}$.
Distinctively, when the frozen valley starts to be probed, 
$\eta$ exhibits a tiny jump for the QAH and ALL states but switches sign for the QVH state
and progressively vanishes with further advancing $\omega'$.
By contrast, the LAF state exhibits little circular dichroism, 
because the two spin flavors have opposite masses $\!\sim\!2$~meV.

\textcolor{blue}{\it{Discussion.---}}
The absorption spectroscopy (or optical conductivity) and circular dichroism provides efficient means 
not only to measure the flavor dependent mass $\lambda^{s\tau}$ 
but also to distinguish the competing ordered states.
It follows that the photoluminescence can also be circularly and valley polarized,
controllable by the external fields. 
Similarly, opto-valley, -spin, and -charge Hall effects may be feasible upon the pumping of these fields, 
given the nontrivial conduction-band Berry curvature~\cite{Zhang2011}.

Three comments are in order. 
(i) Graphene does not reflect a significant amount of light as opposed to TI's,  
and the interaction effects are more pronounced in suspended samples.
Transmission spectroscopy is thus suggested to study the predicted effects.
(ii) In non-equilibrium periodically driven systems, interactions and band populations 
are important and delicate issues for Floquet states~\cite{Bukov2015,DAlessio2015,Seetharam2015}.
In our case, additional terms beyond Eq.~(\ref{UA}) produced by interactions 
can be safely ignored, as we focus on the {\it weak} interaction instability implied by peculiar band structures.
Given the fact that absorbance per graphene layer is only $\pi\alpha\!\sim\!2.3\%$~\cite{ab1,ab2}, 
presumably the band population is almost intact in the presence of the high-frequency light.
(iii) Our Hartree-Fock theory only captures the most essential physics of Floquet bilayer graphene.
The weak yet important interaction parameters $V_z$ and $V_s'$ 
may vary from case to case and should be determined by future experiments.
The phase boundaries in Fig.~\ref{fig-diagram} are likely to be quantitatively modified 
by the thermal proliferation of domain walls separating different ordered states~\cite{Li2014}.

At single-particle level, the Floquet idea~\cite{Oka2009,Kitagawa2010,Morell2012,Piskunow2014} was 
theoretically applied to TI's~\cite{Lindner2011} and semimetals~\cite{Ezawa2013,Chan2016,Yan2016},
and experimentally realized in photonic crystals~\cite{Rechtsman2013}, 
TI surfaces~\cite{Wang2013,Galitski},
and ultracold atoms~\cite{Jotzu2014}.
Our proposal paves the way for generalizing the idea to a paradigmatic many-body system (i.e., chiral few-layer graphene),
and revealed the significant roles played by interactions 
in stabilizing topologically ordered states (e.g. the ALL state) hardly accessible at single-particle level.

\textcolor{blue}{\it{Acknowledgement.---}}
This work is supported by NSF (PHY-1505496), ARO (W911NF-17-1-0128), AFOSR (FA9550-16-1-0387), and UTD Research Enhancement Funds.


\begin{thebibliography}{99}

\bibitem{Neto2009} A. H. Castro Neto, F. Guinea, N. M. R. Peres, K. S. Novoselov, and A. K. Geim, \textit{The electronic properties of graphene}, \href{http://dx.doi.org/10.1103/RevModPhys.81.109}{Rev. Mod. Phys. \textbf{81}, 109 (2009).}

\bibitem{Min-Review} H. Min and A. H. MacDonald, \textit{Electronic structure of multilayer graphene}, \href{http://dx.doi.org/10.1143/PTPS.176.227 }{Prog. Theor. Phys. Suppl. \textbf{176}, 227 (2008).}

\bibitem{Zhang2011} F. Zhang, J. Jung, G. A. Fiete, Q. Niu, and A. H. MacDonald, \textit{Spontaneous quantum Hall states in chirally stacked few-layer graphene systems}, \href{http://dx.doi.org/10.1103/PhysRevLett.106.156801}{Phys. Rev. Lett. \textbf{106}, 156801 (2011)}.

\bibitem{Qiao-Niu} Y. Ren, Z. Qiao, and Q. Niu, \textit{Topological phases in two-dimensional materials: a review}, \href{http://dx.doi.org/10.1088/0034-4885/79/6/066501}{Rep. Prog. Phys. \textbf{79}, 066501 (2016).} 

\bibitem{McCann2006} E. McCann and V. I. Fal'ko, \textit{Landau-level degeneracy and quantum Hall effect in a graphite bilayer}, \href{http://dx.doi.org/10.1103/PhysRevLett.96.086805}{Phys. Rev. Lett. \textbf{96}, 086805 (2006).}

\bibitem{Ohta2006} T. Ohta, A. Bostwick, T. Seyller, K. Horn, and E. Rotenberg, \textit{Controlling the electronic structure of bilayer graphene}, \href{http://dx.doi.org/10.1126/science.1130681}{Science \textbf{313}, 951 (2006)}.

\bibitem{YBZ2009} Y. Zhang, T.-T. Tang, C. Girit, Z. Hao, M. C. Martin, A. Zettl, M. F. Crommie, Y. R. Shen, and F. Wang, \textit{Direct observation of a widely tunable bandgap in bilayer graphene}, \href{http://dx.doi.org/10.1038/nature08105}{Nature \textbf{459}, 820 (2009)}.

\bibitem{Mak2009} K. F. Mak, C. H. Lui, J. Shan, and T. F. Heinz, \textit{Observation of an electric-field-induced band gap in bilayer graphene by infrared spectroscopy}, \href{http://dx.doi.org/10.1103/PhysRevLett.102.256405}{Phys. Rev. Lett. \textbf{102}, 256405 (2009).}

\bibitem{Lui} C. H. Lui, Z. Li, K. F. Mak, E. Cappelluti, and T. F. Heinz, \textit{Observation of an electrically tunable band gap in trilayer graphene}, \href{http://dx.doi.org/10.1038/nphys2102}{Nat. Phys. \textbf{7}, 944 (2011)}.

\bibitem{Zhu} K. Zou, F. Zhang, C. Clapp, A. H. MacDonald, and J. Zhu, \textit{Transport studies of dual-gated ABC and ABA trilayer graphene: band gap opening and band structure tuning in very large perpendicular electric fields}, \href{http://dx.doi.org/10.1021/nl303375a}{Nano Lett. \textbf{13}, 369 (2013)}.

\bibitem{Zhang2015} F. Zhang, \textit{Spontaneous chiral symmetry breaking in bilayer graphene}, \href{http://dx.doi.org/10.1016/j.synthmet.2015.07.028}{Synthetic Metals \textbf{210}, 9 (2015).}

\bibitem{Min2008} H. Min, G. Borghi, M. Polini, and A. H. MacDonald, \textit{Pseudospin magnetism in graphene}, \href{http://dx.doi.org/10.1103/PhysRevB.77.041407}{Phys. Rev. B \textbf{77}, 041407(R) (2008)}.

\bibitem{Sun} K. Sun, H. Yao, E. Fradkin, and S. A. Kivelson, \textit{Topological insulators and nematic phases from spontaneous symmetry breaking in 2D Fermi systems with a quadratic band crossing}, \href{http://dx.doi.org/10.1103/PhysRevLett.103.046811}{Phys. Rev. Lett. \textbf{103}, 046811 (2009)}.

\bibitem{Zhang2010} F. Zhang, H. Min, M. Polini, and A. H. MacDonald, \textit{Spontaneous inversion symmetry breaking in graphene bilayers}, \href{http://dx.doi.org/10.1103/PhysRevB.81.041402}{Phys. Rev. B \textbf{81}, 041402(R) (2010).}

\bibitem{Levitov1} R. Nandkishore and L. Levitov, \textit{Dynamical screening and exotic instability in bilayer graphene}, \href{http://dx.doi.org/10.1103/PhysRevLett.104.156803}{Phys. Rev. Lett. \textbf{104}, 156803 (2010).}

\bibitem{Vafek1} O. Vafek and K. Yang, \textit{Many-body instability of Coulomb interacting bilayer graphene: Renormalization group approach}, \href{http://dx.doi.org/10.1103/PhysRevB.81.041401}{Phys. Rev. B \textbf{81}, 041401(R) (2010).}

\bibitem{Falko1} Y. Lemonik, I. L. Aleiner, C. Toke, and V. I. Fal'ko, \textit{Spontaneous symmetry breaking and Lifshitz transition in bilayer graphene}, \href{http://dx.doi.org/10.1103/PhysRevB.82.201408}{Phys. Rev. B \textbf{82}, 201408(R) (2010).}

\bibitem{Levitov2}  R. Nandkishore and L. Levitov, \textit{Quantum anomalous Hall state in bilayer graphene}, \href{http://dx.doi.org/10.1103/PhysRevB.82.115124}{Phys. Rev. B \textbf{82}, 115124 (2010).}

\bibitem{Jung2011} J. Jung, F. Zhang, and A. H. MacDonald, \textit{Lattice theory of pseudospin ferromagnetism in bilayer graphene: Competing interaction-induced quantum Hall states}, \href{http://dx.doi.org/10.1103/PhysRevB.83.115408}{Phys. Rev. B \textbf{83}, 115408 (2011).}

\bibitem{Zhang2012} F. Zhang and A. H. MacDonald, \textit{Distinguishing spontaneous quantum Hall states in bilayer graphene}, \href{http://dx.doi.org/10.1103/PhysRevLett.108.186804}{Phys. Rev. Lett. \textbf{108}, 186804 (2012)}.

\bibitem{Trushin2011} M. Trushin and J. Schliemann, \textit{Polarization-sensitive absorption of THz radiation by interacting electrons in chirally stacked multilayer graphene}, \href{http://dx.doi.org/10.1088/1367-2630/14/9/095005}{New. J. Phys. \textbf{14}, 095005 (2012)}.

\bibitem{Honerkamp1} T. C. Lang, Z. Y. Meng, M. M. Scherer, S. Uebelacker, F. F. Assaad, A. Muramatsu, C. Honerkamp, and S. Wessel, \textit{Antiferromagnetism in the Hubbard model on the Bernal-stacked honeycomb bilayer}, \href{http://dx.doi.org/10.1103/PhysRevLett.109.126402}{Phys. Rev. Lett. \textbf{109}, 126402 (2012).}

\bibitem{Honerkamp2} M. M. Scherer, S. Uebelacker, and C. Honerkamp, \textit{Instabilities of interacting electrons on the honeycomb bilayer}, \href{http://dx.doi.org/10.1103/PhysRevB.85.235408}{Phys. Rev. B \textbf{85}, 235408 (2012).}

\bibitem{Falko2} Y. Lemonik, I. Aleiner, and V. I. Fal'ko, \textit{Competing nematic, antiferromagnetic, and spin-flux orders in the ground state of bilayer graphene}, \href{http://dx.doi.org/10.1103/PhysRevB.85.245451}{Phys. Rev. B \textbf{85}, 245451 (2012).}

\bibitem{Vafek2} V. Cvetkovic, R. E. Throckmorton, and O. Vafek, \textit{Electronic multicriticality in bilayer graphene}, \href{http://dx.doi.org/10.1103/PhysRevB.86.075467}{Phys. Rev. B \textbf{86}, 075467 (2012).}

\bibitem{Yan} X.-Z. Yan and C. S. Ting, \textit{Absence of gapped broken inversion symmetry phase of electrons in bilayer graphene under the renormalized ring-diagram approximation}, \href{http://dx.doi.org/10.1103/PhysRevB.86.125438}{Phys. Rev. B \textbf{86}, 125438 (2012)}; \textit{Possible broken inversion and time-reversal symmetry state of electrons in bilayer graphene}, \href{http://dx.doi.org/10.1103/PhysRevB.86.235126}{Phys. Rev. B \textbf{86}, 235126 (2012)}; \textit{Ordered-current state of electrons in bilayer graphene}, \href{http://dx.doi.org/10.1103/PhysRevB.88.045410}{Phys. Rev. B \textbf{88}, 045410 (2013).}

\bibitem{Gorbar} E. V. Gorbar, V. P. Gusynin, V. A. Miransky, and I. A. Shovkovy, \textit{Coexistence and competition of nematic and gapped states in bilayer graphene}, \href{http://dx.doi.org/10.1103/PhysRevB.86.125439}{Phys. Rev. B \textbf{86}, 125439 (2012).}

\bibitem{Min12} F. Zhang, H. Min, and A. H. MacDonald, \textit{Competing ordered states in bilayer graphene}, \href{http://dx.doi.org/10.1103/PhysRevB.86.155128}{Phys. Rev. B \textbf{86}, 155128 (2012).}

\bibitem{Kharitonov} M. Kharitonov, \textit{Antiferromagnetic state in bilayer graphene}, \href{http://dx.doi.org/10.1103/PhysRevB.86.195435}{Phys. Rev. B \textbf{86}, 195435 (2012).}

\bibitem{Varma} L. Zhu, V. Aji, and C. M. Varma, \textit{Ordered loop current states in bilayer graphene}, \href{http://dx.doi.org/10.1103/PhysRevB.87.035427}{Phys. Rev. B \textbf{87}, 035427 (2013).}

\bibitem{Li2014} X. Li, F. Zhang, Q. Niu, and A. H. MacDonald, \textit{Spontaneous layer-pseudospin domain walls in bilayer graphene}, \href{http://dx.doi.org/10.1103/PhysRevLett.113.116803}{Phys. Rev. Lett. \textbf{113}, 116803 (2014).}

\bibitem{Rossi} J. Zhang, R. Nandkishore, and E. Rossi, \textit{Disorder-tuned selection of order in bilayer graphene}, \href{http://dx.doi.org/10.1103/PhysRevB.91.205425}{Phys. Rev. B. \textbf{91}, 205425 (2015).}

\bibitem{ex1} J. Martin, B. E. Feldman, R. T. Weitz, M. T. Allen, and A. Yacoby, \textit{Local compressibility measurements of correlated states in suspended bilayer graphene}, \href{http://dx.doi.org/10.1103/PhysRevLett.105.256806}{Phys. Rev. Lett. \textbf{105}, 256806 (2010).}

\bibitem{ex2} R. T. Weitz, M. T. Allen, B. E. Feldman, J. Martin, and A. Yacoby, \textit{Broken-symmetry states in doubly gated suspended bilayer graphene}, \href{http://dx.doi.org/10.1126/science.1194988 }{Science \textbf{330}, 812 (2010).}

\bibitem{ex} A. S. Mayorov, D. C. Elias, M. Mucha-Kruczynski, R. V. Gorbachev, T. Tudorovskiy, A. Zhukov, S. V. Morozov, M. I. Katsnelson, V. I. Fal'ko, A. K. Geim, and K. S. Novoselov, \textit{Interaction-driven spectrum reconstruction in bilayer graphene}, \href{http://dx.doi.org/10.1126/science.1208683 }{Science \textbf{333}, 860 (2011).}

\bibitem{ex3} W. Bao, L. Jing, J. Velasco, Jr., Y. Lee, G. Liu, D. Tran, B. Standley, M. Aykol, S. B. Cronin, D. Smirnov, M. Koshino, E. McCann, M. Bockrath, and C. N. Lau, \textit{Stacking-dependent band gap and quantum transport in trilayer graphene}, \href{http://dx.doi.org/10.1038/nphys2103}{Nat. Phys. \textbf{7}, 948 (2011).}

\bibitem{ex4} F. Freitag, J. Trbovic, M. Weiss, and C. Sch\"onenberger, \textit{Spontaneously gapped ground state in suspended bilayer graphene}, \href{http://dx.doi.org/10.1103/PhysRevLett.108.076602}{Phys. Rev. Lett. \textbf{108}, 076602 (2012).}

\bibitem{ex5} J. Velasco Jr., L. Jing, W. Bao, Y. Lee, P. Kratz, V. Aji, M. Bockrath, C. N. Lau, C. Varma, R. Stillwell, D. Smirnov, F. Zhang, J. Jung, and A. H. MacDonald, \textit{Transport spectroscopy of symmetry-broken insulating states in bilayer graphene}, \href{http://dx.doi.org/10.1038/nnano.2011.251}{Nat. Nanotech. \textbf{7}, 156 (2012)}.

\bibitem{ex6} W. Bao, J. Velasco, Jr., F. Zhang, L. Jing, B. Standley, D. Smirnov, M. Bockrath, A. H. MacDonald, and C. N. Lau, \textit{Evidence for a spontaneous gapped state in ultraclean bilayer graphene}, \href{http://dx.doi.org/10.1073/pnas.1205978109}{Proc. Natl. Acad. Sci. U.S.A. \textbf{109}, 10802 (2012).}

\bibitem{ex7} A. Veligura, H. J. van Elferen, N. Tombros, J. C. Maan, U. Zeitler, and B. J. van Wees, \textit{Transport gap in suspended bilayer graphene at zero magnetic field}, \href{http://dx.doi.org/10.1103/PhysRevB.85.155412}{Phys. Rev. B \textbf{85}, 155412 (2012).}

\bibitem{ex8} F. Freitag, M. Weiss, R. Maurand, J. Trbovic, and C. Sch\"onenberger, \textit{Spin symmetry of the bilayer graphene ground state}, \href{http://dx.doi.org/10.1103/PhysRevB.87.161402}{Phys. Rev. B \textbf{87}, 161402(R) (2013).}

\bibitem{ex9} J. Velasco Jr., Y. Lee, F. Zhang, K. Myhro, D. Tran, M. Deo, D. Smirnov, A. H. MacDonald, and C. N. Lau, \textit{Competing ordered states with filling factor two in bilayer graphene}, \href{http://dx.doi.org/10.1038/ncomms5550}{Nat. Commun. \textbf{5}, 4550 (2014).}

\bibitem{ex10} Y. Lee, D. Tran, K. Myhro, J. Velasco Jr., N. Gillgren, C. N. Lau, Y. Barlas, J. M.
Poumirol, D. Smirnov, and F. Guinea, \textit{Competition between spontaneous symmetry breaking and single-particle gaps in trilayer graphene}, \href{http://dx.doi.org/10.1038/ncomms6656}{Nat. Commun. \textbf{5}, 5656 (2014).}
 
\bibitem{ex11} A. L. Grushina, D.-K. Ki, M. Koshino, A. A. L. Nicolet, C. Faugeras, E. McCann, M.
Potemski, and A. F. Morpurgo, \textit{Insulating state in tetralayers reveals an even-odd interaction effect in multilayer graphene}, \href{http://dx.doi.org/10.1038/ncomms7419}{Nat. Commun. \textbf{6}, 6419 (2015).}      

\bibitem{Haldane1988} F. D. M. Haldane, \textit{Model for a quantum Hall effect without Landau levels: condensed-matter realization of the ``parity anomaly"}, \href{http://dx.doi.org/10.1103/PhysRevLett.61.2015}{Phys. Rev. Lett. \textbf{61}, 2015 (1988).}

\bibitem{Kane2005} C. L. Kane and E. J. Mele, \textit{Quantum spin Hall effect in graphene}, \href{http://dx.doi.org/10.1103/PhysRevLett.95.226801}{Phys. Rev. Lett. \textbf{95}, 226801 (2005).}

\bibitem{Wang2013} Y. H. Wang, H. Steinberg, P. Jarillo-Herrero, and N. Gedik, \textit{Observation of Floquet-Bloch states on the surface of a topological insulator}, \href{http://dx.doi.org/10.1126/science.1239834}{Science \textbf{342}, 453 (2013)}.

\bibitem{Galitski} B. M. Fregoso, Y. H. Wang, N. Gedik, and V. Galitski, \textit{Driven electronic states at the surface of a topological insulator}, \href{http://dx.doi.org/10.1103/PhysRevB.88.155129}{Phys. Rev. B \textbf{88}, 155129 (2013).}

\bibitem{Goldman2015} N. Goldman and J. Dalibard, \textit{Periodically driven quantum systems: effective Hamiltonians and engineered gauge fields}, \href{http://dx.doi.org/10.1103/PhysRevX.4.031027}{Phys. Rev. X \textbf{4}, 031027 (2014)}.

\bibitem{Bukov2015} M. Bukov, L. D'Alessio, and A. Polkovnikov, \textit{Universal high-frequency behavior of periodically driven systems: from dynamical stabilization to Floquet engineering}, \href{http://dx.doi.org/10.1080/00018732.2015.1055918}{Advances in Physics, \textbf{64}, 139 (2015).}

\bibitem{ZhangABC} F. Zhang, B. Sahu, H. Min, and A. H. MacDonald, \textit{Band structure of ABC-stacked graphene trilayers}, \href{http://dx.doi.org/10.1103/PhysRevB.82.035409}{Phys. Rev. B \textbf{82}, 035409 (2010).}

\bibitem{Oka2009} T. Oka and H. Aoki, \textit{Photovoltaic Hall effect in graphene}, \href{http://dx.doi.org/10.1103/PhysRevB.79.081406}{Phys. Rev. B \textbf{79}, 081406(R) (2009).}

\bibitem{Kitagawa2010} T. Kitagawa, E. Berg, M. Rudner, and E. Demler, \textit{Topological characterization of periodically driven quantum systems}, \href{http://dx.doi.org/10.1103/PhysRevB.82.235114}{Phys. Rev. B \textbf{82}, 235114 (2010).}

\bibitem{Morell2012} E. S. Morell and L. E. F. Foa Torres, \textit{Radiation effects on the electronic properties of bilayer graphene}, \href{https://doi.org/10.1103/PhysRevB.86.125449}{Phys. Rev. B \textbf{86}, 125449 (2012).}

\bibitem{Piskunow2014} P. M. Perez-Piskunow, Gonzalo Usaj, C. A. Balseiro, and L. E. F. Foa Torres, \textit{Floquet chiral edge states in graphene}, \href{https://doi.org/10.1103/PhysRevB.89.121401}{Phys. Rev. B \textbf{89}, 121401 (2014).}

\bibitem{vc1} I. Martin, Y. M. Blanter, and A. F. Morpurgo, \textit{Topological confinement in bilayer graphene}, \href{http://dx.doi.org/10.1103/PhysRevLett.100.036804}{Phys. Rev. Lett. \textbf{100}, 036804 (2008).}

\bibitem{vc2} J. Li, I. Martin, M. B\"uttiker, and A. F. Morpurgo, \textit{Topological origin of subgap conductance in insulating bilayer graphene}, \href{http://dx.doi.org/10.1038/nphys1822}{Nat. Phys. \textbf{7}, 38 (2011).}

\bibitem{vc3} Z. Qiao, J. Jung, Q. Niu, and A. H. MacDonald, \textit{Electronic highways in bilayer graphene}, \href{http://dx.doi.org/10.1021/nl201941f}{Nano Lett. \textbf{11}, 3453 (2011).}

\bibitem{vc4} M. Zarenia, J. M. Pereira, Jr., G. A. Farias, and F. M. Peeters, \textit{Chiral states in bilayer graphene: magnetic field dependence and gap opening}, \href{http://dx.doi.org/10.1103/PhysRevB.84.125451}{Phys. Rev. B \textbf{84}, 125451 (2011).}


\bibitem{vc5} F. Zhang, A. H. MacDonald, and E. J. Mele, \textit{Valley Chern numbers and boundary modes in gapped bilayer graphene}, \href{http://dx.doi.org/10.1073/pnas.1308853110}{Proc. Natl. Acad. Sci. USA \textbf{110}, 10546 (2013).}


\bibitem{vc6} A. Vaezi, Y. Liang, D. H. Ngai, L. Yang, and E.-A. Kim, \textit{Topological edge states at a tilt boundary in gated multilayer graphene}, \href{http://dx.doi.org/10.1103/PhysRevX.3.021018}{Phys. Rev. X \textbf{3}, 021018 (2013).}


\bibitem{vc7} L. Ju, Z. Shi, N. Nair, Y. Lv, C. Jin, J. Velasco, Jr., C. Ojeda-Aristizabal, H. A. Bechtel, M. C. Martin, A. Zettl, J. Analytis, and F. Wang, \textit{Gate-controlled topological conducting channels in bilayer graphene}, \href{http://dx.doi.org/10.1038/nature14364}{Nature \textbf{520}, 650 (2015)}.

\bibitem{vc9}J. Li, K. Wang, K. J. McFaul, Z. Zern, Y. F. Ren, K. Watanabe, T. Taniguchi, Z. H. Qiao, and J. Zhu, \textit{Experimental observation of edge states at the line junction of two oppositely biased bilayer graphene}, \href{http://dx.doi.org/10.1038/nnano.2016.158}{Nature Nanotechnology \textbf{11}, 1060 (2016).}


\bibitem{vc10} L.-J. Yin, H. Jiang, J.-B. Qiao, and L. He, \textit{Direct imaging of topological edge states at a bilayer graphene domain wall}, \href{http://dx.doi.org/10.1038/ncomms11760}{Nat. Commun. \textbf{7}, 11760 (2016).}

\bibitem{vc11} L. Jiang, Z. Shi, B. Zeng, S. Wang, J.-H. Kang, T. Joshi, C. Jin, L. Ju, J. Kim, T. Lyu, Y.-R. Shen, M. Crommie, H.-J. Gao, and F. Wang, \textit{Soliton-dependent plasmon reflection at bilayer graphene domain walls}, \href{http://dx.doi.org/10.1038/nmat4653}{Nature Materials \textbf{15}, 840 (2016).}.

\bibitem{ab1} R. R. Nair, P. Blake, A. N. Grigorenko, K. S. Novoselov, T. J. Booth, T. Stauber, N. M. R. Peres, and A. K. Geim, \textit{Fine structure constant defines visual transparency of graphene}, \href{http://dx.doi.org/10.1126/science.1156965 }{Science \textbf{320}, 1308 (2008).}

\bibitem{ab2} K. F. Mak, M. Y. Sfeir, Y. Wu, C. H. Lui, J. A. Misewich, and T. F. Heinz, \href{http://dx.doi.org/10.1103/PhysRevLett.101.196405}{Phys. Rev. Lett. \textbf{101}, 196405 (2008).}

\bibitem{Probe1} X. Xu, W. Yao, D. Xiao, and T. F. Heinz, \textit{Spin and pseudospins in layered transition metal dichalcogenides}, \href{http://dx.doi.org/10.1038/nphys2942}{Nat. Phys. \textbf{10}, 343 (2014).}

\bibitem{Probe2} W. Yao, D. Xiao, and Q. Niu, \textit{Valley-dependent optoelectronics from inversion symmetry breaking}, \href{http://dx.doi.org/10.1103/PhysRevB.77.235406}{Phys. Rev. B \textbf{77}, 235406 (2008)}.

\bibitem{Probe3} M. Trushin and J. Schliemann, \textit{Pseudospin in optical and transport properties of graphene}, \href{http://dx.doi.org/10.1103/PhysRevLett.107.156801}{Phys. Rev. Lett. \textbf{107}, 156801 (2011)}.

\bibitem{Probe4} H. Pan, X. Li, F. Zhang, and S. A. Yang, \textit{Perfect valley filter in a topological domain wall}, \href{http://dx.doi.org/10.1103/PhysRevB.92.041404}{Phys. Rev. B \textbf{92}, 041404(R) (2015)}.

\bibitem{DAlessio2015} L. D'Alessio and M. Rigol, \textit{Dynamical preparation of Floquet Chern insulators}, \href{http://dx.doi.org/10.1038/ncomms9336}{Nat. Commun. \textbf{6}, 8336 (2015).}

\bibitem{Seetharam2015} K. I. Seetharam, C.-E. Bardyn, N. H. Lindner, M. S. Rudner, and G. Refael, \textit{Controlled population of Floquet-Bloch state via coupling to Bose and Fermi baths}, \href{http://dx.doi.org/10.1103/PhysRevX.5.041050}{Phys. Rev. X \textbf{5}, 041050 (2015).}

\bibitem{Lindner2011} N. H. Lindner, G. Refael, and V. Galitski, \textit{Floquet topological insulator in semiconductor quantum wells}, \href{http://dx.doi.org/10.1038/nphys1926}{Nat. Phys. \textbf{7}, 490 (2011).}

\bibitem{Ezawa2013} M. Ezawa, \textit{Photoinduced topological phase transition and a single Dirac-cone state in silicene}, \href{http://dx.doi.org/10.1103/PhysRevLett.110.026603}{Phys. Rev. Lett. \textbf{110}, 026603 (2013).}

\bibitem{Chan2016} C. Chan, P. A. Lee, K. S. Burch, J. H. Han, and Y. Ran, \textit{When chiral photons meet chiral fermions: photoinduced anomalous Hall effects in Weyl semimetals}, \href{http://dx.doi.org/10.1103/PhysRevLett.116.026805}{Phys. Rev. Lett. \textbf{116}, 026805 (2016).}

\bibitem{Yan2016} Z. Yan and Z. Wang, \textit{ Tunable Weyl semimetals in periodically driven nodal line semimetals}, \href{http://dx.doi.org/10.1103/PhysRevLett.117.087402}{Phys. Rev. Lett. \textbf{117}, 087402 (2016).}

\bibitem{Rechtsman2013} M. C. Rechtsman, J. M. Zeuner, Y. Plotnik, Y. Lumer, D. Podolsky, F. Dreisow, S. Nolte, M. Segev, and A. Szameit, \textit{Photonic Floquet topological insulators}, \href{http://dx.doi.org/10.1038/nature12066}{Nature \textbf{496}, 196 (2013).}

\bibitem{Jotzu2014} G. Jotzu, M. Messer, R. Desbuquois, M. Lebrat, T. Uehlinger, D. Greif, and T. Esslinger, \textit{Experimental realization of the topological Haldane model with ultracold fermions}, \href{http://dx.doi.org/10.1038/nature13915}{Nature \textbf{515}, 237 (2014).}

\end{thebibliography}
\end{document}